\input{aipcheck}

\documentclass[
    ,final            
  ]
  {aipproc}

\layoutstyle{6x9}

\begin{document}

\title{Presupernova evolution and explosion of massive stars with mass loss}


\author{M. Limongi}{
  address={INAF - Osservatorio Astronomico di Roma, Via Frascati 33, I-00040, Rome, Italy}
  ,altaddress={Department of Astronomy, University of Tokyo, Bunkyo-ku, Tokyo 113-0033, Japan}
}

\author{A. Chieffi}{
  address={INAF - Istituto di Astrofisica Spaziale e Fisica Cosmica, Via Fosso del Cavaliere, I-00133, Rome, Italy}
  ,altaddress={Centre for Stellar and Planetary Astrophysics, School of Mathematical Sciences, P.O. Box 
               28M, Monash University, Victoria 3800, Australia}
}


\begin{abstract}

We review the main properties of solar metallicity massive stars in the range 11-120 $\rm M_\odot$. The influence of the mass loss 
on the hydrostatic burning stages as well as the final explosion is discussed in some detail. We find that the minimum masses that 
enter the WNL, WNE and WC stages are 30 $\rm M_\odot$, 35 $\rm M_\odot$ and 40 $\rm M_\odot$ respectively; the limiting mass between 
stars exploding as SNII and SNIb/c is between 30 and 35 $\rm M_\odot$; the limiting mass between stars forming neutron stars and 
black holes after the explosion is between 25-30 $\rm M_\odot$. We also discuss the properties of the chemical yields integrated 
over a Salpeter IMF and we find that stars with $\rm M\ge 35~M_\odot$ contribute for $\sim 60\%$ to the production of C, N and for 
$\sim 40\%$ to the production Sc and s-process elements up to Zr, while they do not produce any intermediate mass element because of 
the large remnant masses. \end{abstract}

\maketitle

\section{Introduction}

Massive stars, i.e. the stars that evolve through all the hydrostatic nuclear burning stages in a quiescent way up to the explosion, 
play a key role in many astrophysical fields. They light up regions of stellar birth, may induce star formation, produce most of the 
elements, especially those necessary to life, are responsible for the mixing of the interstellar medium and also contribute to the 
production of neutron stars and black holes. For all these reasons they play a pivotal role in the evolution of the galaxies. 
Massive stars also produce some long-lived radioactive isotopes like, e.g., $\rm ^{26}Al$,$\rm ^{44}Ti$, $\rm ^{56}Co$,$\rm 
^{60}Fe$, whose radioactive transitions may give rise to $\gamma$-ray photons that can be detected by sufficiently sensitive 
instruments presently in space (INTEGRAL, RHESSI). The measurements of the $\gamma$-ray signals from the decay of species with 
lifetimes shorter than those typical of the evolution of the Galaxy clearly demonstrate that nucleosynthesis is still active today. 
Last, but no least, the collapse of massive stars is likely responsible to the processes leading to some kind of Gamma Ray Burst 
events. Massive stars are of such astrophysical relevance that a proper understanding of their evolution and explosion is mandatory.

In this review we will briefly discuss the main evolutionary properties of almost the full range of massive stars (i.e. between 11 
and 120 $\rm M_\odot$) of initial solar composition during all the hydrostatic burning stages from the main sequence phase up to the 
onset of the iron core collapse. Special attention will be payed to the effects of mass loss. We will also discuss the basic 
properties of their explosive yields together to their contribution to the chemical evolution of the Galaxy.

\section{The stellar models}

All the models presented in this paper have been computed by means of the latest version of the FRANEC that is described in detail 
in \cite{LC06}. Let us just mention here that mass loss is included following \cite{Vink2000} for the blue supergiant phase, and 
\cite{Deja1989} for the red supergiant phase. For the Wolf-Rayet phase the mass loss rate provided by \cite{NL2000} has been 
adopted. The models have initial solar composition \cite{AG1989} and range in mass between 11 and 120 $\rm M_\odot$.

\section{H and He burning}

The core H burning is the first nuclear burning stage and, in these stars, 
is powered by the CNO cycle. The strong dependence of this cycle on the 
temperature implies the presence of a convective core that reaches its 
maximum extension just at the beginning of the central H burning and that 
then recedes in mass as the central H is burnt. Mass loss is quite 
efficient during this phase and increases substantially with the luminosity 
of the star (i.e. the initial mass). It leads to a significant reduction of 
the total mass of the star in stars initially more massive than $\rm \sim 
40~M_\odot$. For example, the $\rm 60~M_\odot$ model has a total mass at 
core H exhaustion equal to $\rm 38~M_\odot$ while the $\rm 120~M_\odot$ 
ends the H burning as a $\rm 56~M_\odot$. The mass loss rate in this phase
typically ranges from $\rm 10^{-6}~M_\odot$/yr for the 
40 $\rm M_\odot$ to $\rm 10^{- 5}~M_\odot$/yr for 
the 120 $\rm M_\odot$. Most of the massive stars spend a significant 
fraction of their central H burning lifetime as O-type stars (i.e. have 
effective temperatures ranging between 50000 K<$\rm T_{eff}$<33000 K) 
and the minimum mass that becomes an O-type star is the $\rm 
14~M_\odot$ in our set of models (Figure 1). The fraction of H burning lifetime spent 
by a star as an O-type star increases with the initial mass and varies from 
$\sim 15\%$ for the $\rm 14~M_\odot$ to $\sim 80\%$ for the $\rm 
120~M_\odot$. The H exhausted core (He core) that forms at the central H 
exhaustion scales (in mass size) directly with the initial mass. It is 
worth noting here that the presently adopted mass loss rates in the blue 
supergiant phase do not alter too much the initial mass-He core mass at the 
core H exhaustion. For example, the He core mass of a 80 $\rm M_\odot$ 
model at core H exhaustion without mass loss is 26 $\rm M_\odot$ while for 
the same model with mass loss it reduces to 22 $\rm M_\odot$, i.e., only by 
$\sim 5\%$, although the total mass is reduced to 47 $\rm M_\odot$, i.e., 
by $\sim 40\%$. On the contrary the size of the H convective core, which in 
turn may be affected by the overshooting and/or semiconvection, has a 
strong impact on the initial mass-He core mass relation and hence 
constitutes the greatest uncertainty in the computation of the core H 
burning phase.

\begin{figure}
  \includegraphics[height=.3\textheight]{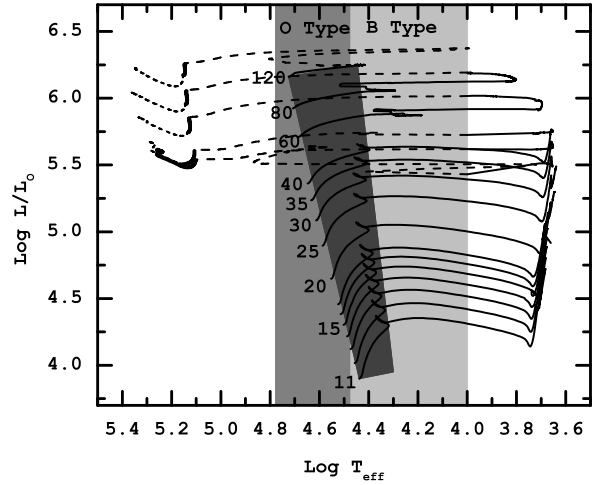}
  \caption{Evolutionary paths of the stellar models in the HR diagram.
The $dark-grey$ shaded area marks the location of core H burning models.
The $dashed$, $thick-solid$ and $dotted$ lines refer to the
WNL, WNE and WC stages respectively. Also shown are the regions
corresponding to the O-type and B-type stars.}
\end{figure}

At the core H exhaustion all the models move toward the red side of the HR 
diagram while the centre contracts until the core He burning begins. The 
further fate of the models is largely driven by the competition between the 
efficiency of mass loss in reducing the H rich envelope during the RSG 
phase and the core He burning timescale. Note that, as soon as the star 
expands and cools, the mass loss rate switches from that provided by 
\cite{Vink2000} to that given by \cite{Deja1989}. Stars initially less 
massive than 30 $\rm M_\odot$ do not loose most of their H rich mantle 
hence they will live up to 
the onset of the core collapse and explode as red supergiants.
Vice versa stars initially more massive than 
this threshold value loose enough mass to end their life as WR stars. 
However, also these stars may spend a fraction of their He burning lifetime 
as red supergiants before becoming WR stars. The fraction of core He 
burning lifetime spent by one of these stars as a RSG ($\rm 
\tau_{red}/\tau_{He}$) obviously depends on the efficiency of the mass loss 
and since it increases significantly with the luminosity (that in turn 
depends on the initial mass of the star), the larger the mass the higher 
the mass loss rate and the smaller the percentage of time spend by a star 
as a red supergiant. For example the 35 $\rm M_\odot$ becomes a WNL star 
when the central He mass fraction is 0.52 and $\rm \tau_{red}/\tau_{He}\sim 
0.38$, while in the 80 $\rm M_\odot$ the transition from RSG to BSG occurs 
when the central He mass fraction is 0.95 and hence $\rm 
\tau_{red}/\tau_{He}$ is less the 0.002. The 120 $\rm M_\odot$ never moves 
to the red because it becomes a WNL already during the latest stages of 
core H burning. It is worth noting here that when a star becomes a Wolf-
Rayet the mass loss rate typically reduces by about one order of magnitude 
compared to that in the RSG phase \citep{NL2000}. When the surface H mass 
fraction abundance drops below roughly 0.4 the star quickly moves to the 
blue side of the HR diagram and continues the core He burning as a WR star. 
As it is well known, central He burning occurs in a convective core whose 
mass size either advances in mass or remains fixed (at most). Such a 
behavior, typical in stars in which the He core mass does not shrink in 
mass during the central He burning, changes in the models in which the He 
core mass reduces during the central He burning (i.e. the WNE/WC/WO stars). 
The stars that in our set of models become at least WNE WR stars (i.e., the 
ones having $\rm X_{sup}<10^{- 5}$ and $\rm (C/N)_{sup}<0.1$ according to 
the standard definition) are those with $\rm M\geq35~M_\odot$. Once the 
full H rich mantle is lost, the further evolution of the stellar model 
depends on the actual He core mass. In fact, as the He core is reduced 
by mass loss the star feels this reduction and tends to behave like a star 
of a smaller mass (i.e., a star having the same actual He core mass). 
Hence, the reduction of the He core (i.e. the total mass) during the core 
He burning has the following effects: 1) the He convective core reduces 
progressively in mass, 2) the He burning lifetime increases, 3) the 
luminosity progressively decreases, 4) the $\rm ^{12}C$ mass fraction at 
core He exhaustion is higher than it would be without mass loss and 4) the 
CO core at the core He exhaustion resemble that of other stars having 
similar He core masses independently on the initial mass of the star.

\begin{figure}
  \includegraphics[height=.3\textheight]{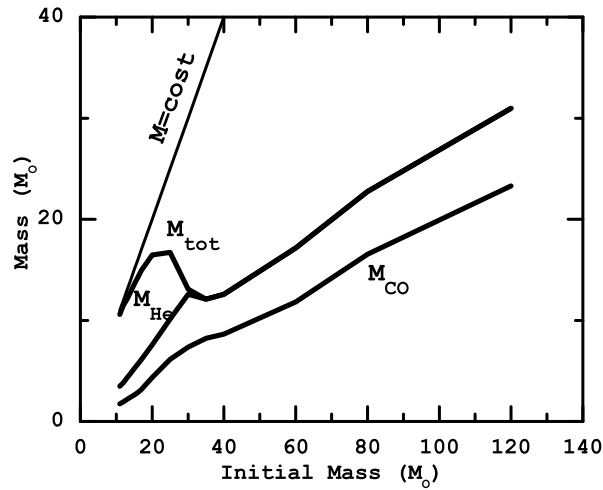}
  \caption{Total mass, He core mass and CO core mass as a function of the
initial mass at core He exhaustion.}
\end{figure}

Stars with mass $\rm M\geq40~M_\odot$ (in our scenario) experience a mass loss strong enough that the total mass is decreased down 
to the mass coordinate marking the maximum extension of the He convective core. Once this happens, the star becomes a WC Wolf-Rayet 
star (i.e., when $\rm (C/N)_{sup}>10$ according to the standard definition) and the products of core He burning are exposed to the 
surface and ejected into the interstellar medium through their stellar wind.

Figure 2 shows the He and the CO core masses at core He exhaustion as a function of the initial mass. It is worth noting that, 
independently on the total amount of mass lost, the CO core shows a clear one to one trend with the mass in the whole range of 
masses. Before closing this section it is worth reminding that obviously the mass loss during both the RSG and the WR phases has a 
strong impact on the WR lifetime and on the specific lifetimes in the various WR stages. Let us recall here few basic rules: 1) the 
higher the mass loss during the RSG phase the earlier the stage (the higher the central He mass fraction) at which the star enters 
the WR phase; 2) the higher the amount of mass lost during the WR phase the lower the actual size of the He core, the longer the He 
burning lifetimes and hence the longer the WR lifetime.

\section{The advanced burning stages}

The nuclear burning stages that follow the central He exhaustion up to the 
presupernova stage have been extensively discussed in the literature 
\cite{WW1995,LSC2000,UN2002,LC2003,CL2004} and hence here we just recall 
few basic properties. The advanced burning evolutionary stages are 
characterized by a strong neutrino emission, the neutrinos being mainly 
produced by pairs production in the center of the star. From the central C 
burning onward, the neutrino luminosity progressively increases and 
supersedes the photon luminosity \cite{LSC2000} by several orders of 
magnitude (up to $\sim 8$) and hence the advanced nuclear burnings must 
speed up enormously in order to counterbalance such dramatic energy losses. 
This also means that the region outside the CO core mass remains 
essentially freezed out during the advanced burning phases up to the moment 
of the explosion.

The evolution of the star during the advanced burning stages is mainly 
driven by the size and the composition ($\rm ^{12}C/^{16}O$) of the CO 
core. In general each burning stage, from the C burning up to the Si 
burning, occurs first at the center and then, as the fuel is completely 
burnt, it shifts in a shell. The nuclear burning shell, then, may induce 
the formation of one of more successive convective zones, that can 
partially overlap. The general trend is that the number of convective shells 
scales inversely with the mass size of the CO core. For example in the 11 
$\rm M_\odot$ model, four consecutive C convective shells form while in the 
60 $\rm M_\odot$ model only one. The complex interplay among burning shells 
and convective zones determines, in a direct way, not only the chemical 
composition but also the final mass-radius relation inside the CO core at 
the presupernova stage. In general, the more efficient the shell burning 
is, the smaller the number of convective shells, the slower is the 
contraction and the shallower is the final mass-radius relation, i.e., the 
higher is the mass of the star the more compact is the CO core.

\begin{figure}
  \includegraphics[height=.3\textheight]{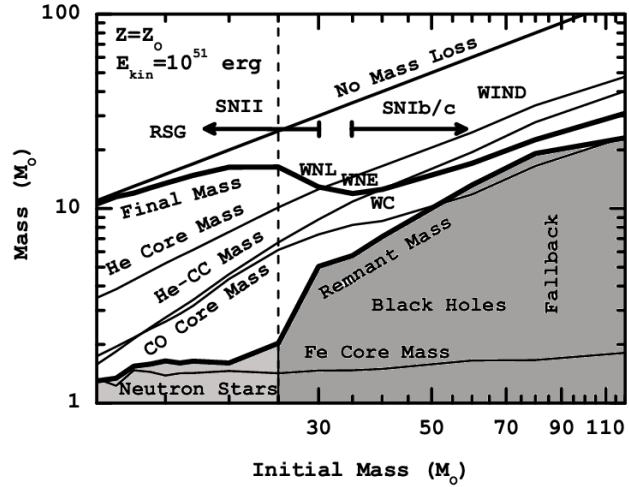}
  \caption{Final masses and remnant masses ($thick-solid$ lines) 
as a function of the initial mass by assuming that all the
models have $\rm E_{kin}=1~foe$ at infinity.
Also shown are the maximum He core, He convective
core and CO core masses, the minimum masses that enter the various
stages of WR, the limiting mass between stars exploding as SNII
and SNIb/c, the limiting mass between stars forming neutron stars
and black holes after the explosion. }
\end{figure}

\section{Simulated explosion and explosive nucleosynthesis}

The chemical composition left by the hydrostatic evolution is partially 
modified by the explosion, especially that of the more internal zones. At 
present there is no self consistent hydrodynamical model for core collapse 
supernovae and consequently we are forced to simulate the explosion in some 
way in order to compute the explosive yields. The idea is to deposit a 
given amount of energy at the base of the exploding envelope and follow the 
propagation of the shock wave that forms by means of a hydro code. The 
initial amount of energy is fixed by requiring a given amount of kinetic 
energy at the infinity (typically of the order of $\rm 10^{51}~erg=1~foe$). 
Whichever is the technique adopted to deposit the energy into the 
presupernova model (piston, kinetic bomb or thermal bomb), in general the 
result is that some amount of material (the innermost one) will fall back 
onto the compact remnant while most of the envelope will be ejected with 
the desired final kinetic energy. The mass separation between remnant and 
ejecta is always referred to as the mass cut. The propagation of the shock 
wave into the exploding envelope induces compression and local heating and 
hence explosive nucleosynthesis. Zones heated up to different peak 
temperatures undergo different kind of explosive nucleosynthesis: complete 
explosive Si burning (for $\rm T_{peak}>5\cdot 10^{9}~K$), incomplete 
explosive Si burning (for $\rm 4\cdot 10^{9}~K<T_{peak}<5\cdot 10^{9}~K$), 
explosive O burning (for $\rm 3.3\cdot 10^{9}~K<T_{peak}<4\cdot 10^{9}~K$), 
explosive Ne burning (for $\rm 2.1\cdot 10^{9}~K<T_{peak}<3.3\cdot 
10^{9}~K$) and explosive C burning (for $\rm 1.9\cdot 
10^{9}~K<T_{peak}<2.1\cdot 10^{9}~K$). Matter heated up to temperatures 
lower than, say, $\rm T_{peak}=1.9\cdot 10^{9}~K$ is not affected by the 
explosion due to the very short explosion timescales. Since the matter 
behind the shock is radiation dominated, the peak temperature at which each 
zone is heated up during the explosion is given by $\rm T_{peak}=(3E/(4\pi 
a R_{PN}^{3}))^{1/4}$, where $\rm R_{PN}$ is the presupernova radius. By 
means of this equation, at each explosive burning corresponds a given 
volume of the star. As a consequence, the mass-radius relation at the 
presupernova stage determines the amount of mass exposed to each explosive 
burning. Following this rule, the more massive stars (i.e., those more 
compact) will produce much more heavy elements (i.e., those produced by the 
explosive burnings) than the less massive ones. However, the steeper is the 
mass-radius relation the higher is the binding energy and hence the larger 
will be, in general, the mass falling back onto the compact remnant. Figure 
3 shows the initial mass-final mass relation in the assumption that the 
ejecta have 1 foe of kinetic energy at the infinity in all the models - the 
explosions have been carried out in the framework of the kinetic bomb. This 
choice (i.e., $\rm E_{kin}=1~foe$ for all the models) implies that in stars 
with masses above 25-30 $\rm M_\odot$ all the CO core, or a great fraction 
of it, fall back onto the compact remnant. As a consequence these stars 
would not eject any product of the explosive burnings as well as those of 
the C convective shell and will leave, after the explosion, black holes 
with masses ranging between 3 and 11 $\rm M_\odot$. In figure 3 there are 
also shown the limiting masses that enter the various WR stages, i.e., WNL 
(30 $\rm M_\odot$), WNE (35 $\rm M_\odot$) and WC (40 $\rm M_\odot$), as 
well as the limiting mass (30- 35 $\rm M_\odot$) between stars exploding as 
Type II SNe and those exploding as Type Ib/c supernovae. These results, 
however, must be taken with caution because they depends on many 
uncertainties related to the both the hydrostatic evolution (mainly mass 
loss) and the explosion (mainly the final kinetic energy at the infinity). 
How this picture changes by changing the mass loss and the final kinetic 
energy will be discussed in a forthcoming paper.

\begin{figure}
  \includegraphics[height=.2\textheight]{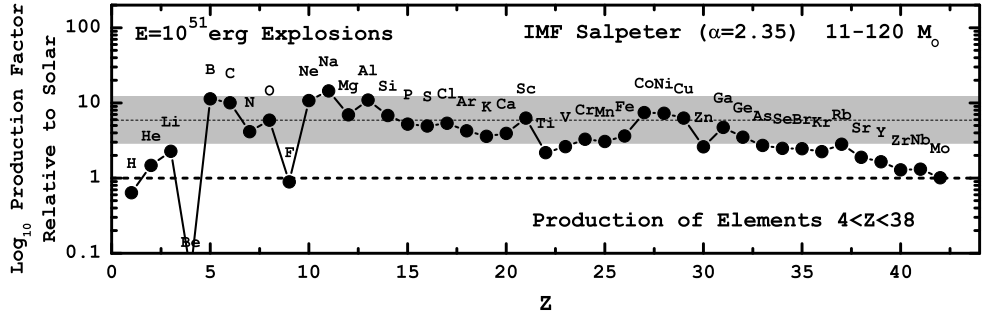}
  \caption{Production factors of a generation of massive stars in the
range 11-120 $\rm M_\odot$ integrated over a Salpeter IMF. All
the models have $\rm E_{kin}=1~foe$ at infinity.}
\end{figure}

\section{Integrated chemical yields}

The first products of the calculations described above are the yields of 
the various isotopes, i.e., the amount of mass of each isotope ejected by 
each star in the interstellar medium. The integration of these yields over 
an initial mass function (IMF) provide the chemical composition of the 
ejecta of a generation of massive stars. Figure 4 shows the production 
factors of all the elements obtained by assuming a Salpeter IMF 
($dn/dm=km^{-2.35}$) provided by a generation of massive stars in the range 
11-120 $\rm M_\odot$. This figure clearly shows that these stars are 
responsible for producing all the elements with 4<Z<38. Moreover, the 
majority of the elements preserve a scaled solar distribution relative to 
O. Other sources, like, e.g. AGB stars and Type Ia SNe, must contribute to 
all the elements that are under produced by massive stars (e.g., the iron 
peak elements and the s-process elements above Ge). The contribution of the 
more massive stars, e.g., stars with $\rm M\geq 35~M_\odot$, to the total 
yield of each element is shown in Figure 5. These more massive stars 
produce roughly $\sim 60\%$ of the total yields of C and N and about $\sim 
40\%$ of Sc and s-process elements. This is the result of the strong mass 
loss experienced by these stars that allows the ejection of these elements, 
synthesized during H (N) and He burning (C, Sc and s-process elements), 
before their destruction during the more advanced burning stages and/or 
during the explosion. On the contrary, these stars do not contribute at all 
to the production of all the intermediate mass and iron peak elements 
because of their large remnant masses.

\begin{figure}
  \includegraphics[height=.2\textheight]{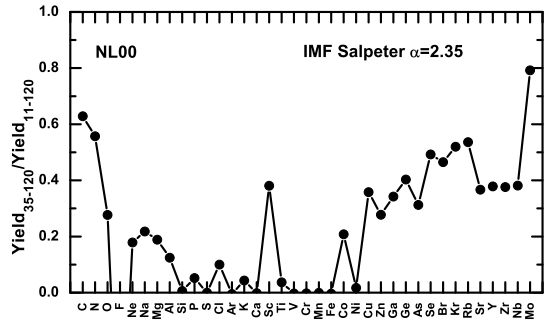}
  \caption{Contribution of stars with $\rm M\geq 35~M_\odot$
to the total yields of a generation of massive stars in the range
11-120 $\rm M_\odot$ integrated over a Salpeter IMF. All
the models have $\rm E_{kin}=1~foe$ at infinity.
}
\end{figure}

\section{Summary and Conclusions}

In this paper we briefly reviewed the main evolutionary properties of 
massive stars during both hydrostatic and explosive stages. The role of 
mass loss has been discussed in some detail. We found the following 
results: 1) stars with $\rm M\leq 25~M_\odot$ explode as RSG while stars 
above this limit explode as BSG; 2) the minimum masses the enter the WNL, 
WNE and WC stages are 30 $\rm M_\odot$, 35 $\rm M_\odot$ and 40 $\rm 
M_\odot$ respectively; 3) the limiting mass between stars exploding as SNII 
and SNIb/c is between 30 and 35 $\rm M_\odot$ and implies a SNIb/c/SNII 
ratio of 0.22 when a Salpeter IMF is adopted; 4) the limiting mass between 
stars forming neutron stars and black holes after the explosion is between 
25-30 $\rm M_\odot$; 5) assuming a Salpeter IMF, stars with $\rm M\geq 
35~M_\odot$ contribute for $\sim 60\%$ to the production of C and N, and for 
$\sim 40\%$ to the production Sc and s- process elements up to Zr, while 
they do not produce any intermediate mass element because of the large 
remnant masses.

As a final comment we want to remark that these results are extremely 
sensitive to many uncertainties among which we emphasize the mass loss and 
the explosion energy at the infinity. A closer investigation on how these 
uncertainties affect the picture presented here will be addressed in a 
forthcoming paper.

\begin{theacknowledgments}

I (ML) would like to thank Ken'ichi Nomoto for many stimulating and helpful discussions and for his very kind and generous 
hospitality at The University of Tokyo during my visit. This work has been supported in part by 21st Century COE Program (QUEST) 
from the MEXT of Japan and in part by the Italian Ministry of the Education, University and Research (MIUR) through the grant PRIN-
2004 (Prot. 2004029938).

\end{theacknowledgments}

\end{document}